\documentclass[pra,showpacs,showkeys]{revtex4} \usepackage{latexsym}
\usepackage{amsfonts} \usepackage{amsmath,revsymb} 

\newcommand{\kt}[1]{\ensuremath{|#1\rangle}}
\newcommand{\br}[1]{\ensuremath {\langle #1|}}
\newcommand{\bk}[2]{\ensuremath {\langle #1|#2 \rangle}}
\newcommand{\kb}[2]{\ensuremath {| #1 \rangle\langle #2|}}

\newcommand{\SU}{{\rm SU}}

\newcommand{\HS}{{\mathcal{H}}}

\begin{document}

\title{Limits on entanglement in rotationally-invariant scattering of spin systems}

\author{N.L.~Harshman\footnote{Electronic address: harshman@american.edu}}
\affiliation{Department of Computer Science, Audio Technology and Physics\\
4400 Massachusetts Ave., NW \\ American University\\ Washington, DC 20016-8058}

\begin{abstract}

This paper investigates the dynamical generation of entanglement in scattering systems, in particular two spin systems that interact via rotationally-invariant scattering.  The spin degrees of freedom of the in-states are assumed to be in unentangled, pure states, as defined by the entropy of entanglement.  Because of the restriction of rotationally-symmetric interactions, perfectly-entangling S-matrices, i.e. those that lead to a maximally entangled out-state, only exist for a certain class of separable in-states.  Using Clebsch-Gordan coefficients for the rotation group, the scattering phases that determine the S-matrix are determined for the case of spin systems with $\sigma = 1/2$, $1$, and $3/2$.
\end{abstract}

\maketitle

\section{Introduction}

The study of entanglement is of central importance to quantum information theory.  Entanglement is the resource for many of its proposed applications, such as quantum computation~\cite{shor} and quantum teleportation~\cite{teleport}.  A variety of experimental procedures for generating entangled quantum systems have been demonstrated and/or proposed and the search continues for new methods that will be stable, scalable, and efficient.  To guide and explain this work, it is important to have theoretical studies of how entanglement is generated in quantum states by the application of dynamical operators.

This work investigates the dynamical generation of entanglement in scattering spin systems by considering the S-matrix operator and how global symmetry properties restrict its form.
This is an complement and alternative to two related methods for investigating the generation of entanglement by dynamics.  One can consider the general properties of global unitary operators by looking at classes of unitary operators in the state space of the quantum system of interest, typically the tensor product of qubits.  See \cite{zhang02} for work related to the approach of this article and \cite{nielsen} for a comprehensive review and bibliography.  In another approach, one can investigate possible interaction Hamiltonians and their exponentiations.  These Hamiltonians could be purely internal or could incorporate the action of outside fields.  The papers \cite{duer01} and \cite{hammerer02} provide good examples of this approach, and many more examples exist which are specific to particular experimental configuration.

The approach of this article specifically explores the generation of entanglement in the non-relativistic, elastic scattering of two distinguishable particles by a central force interaction.  Non-relativistic two-body interactions dominate the dynamics of a gas of trapped ultra-cold atoms, for example, and applications of quantum information theory to that system~\cite{jaksch,brennen,dudarev} require an understanding of dynamical entanglement by scattering~\cite{law}.  More generally, these results apply to any two particle ``scattering-like'' experiment, i.e., a bi-partite spin system where a spherically-symmetric interaction between the two spins can be turned on and off (see example in \cite{cohentdl}).  Systems that are asymptotically non-interacting can be cast into the form of a scattering problem and treated with the techniques below.  Such a sequence could be arranged via controlled interactions, but also appears naturally in the case of finite-range interactions.

In general, the description of entanglement in scattering systems requires that one considers entangled states of with continuous degrees of freedom (see  \cite{kurizki05} and references therein, and the review~\cite{plenioeisert}).  However, for non-relativistic particles with central interactions, there is no mixing between orbital and intrinsic angular momentum~\cite{gw}.  Then the S-matrix can be decomposed into partial waves $S_{s\ell}(\mathbf{p}, E)$ labeled by total spin $s$ and by orbital angular momentum $\ell$ and the S-matrix is diagonal in total momentum $\mathbf{p}$ and energy $E$.  The scattering dynamics do not mix partial waves with different values of $s$ and $l$, and so within each partial wave of orbital angular momentum the entanglement of the spin degrees of freedom is separable from the translational degrees~\cite{harshman_05a}.  As a result, this article only considers entanglement in the spin degrees of freedom.  One cannot make this separation between the translational and rotational degrees of freedom if the interactions are non-central~\cite{gw} or if the system is relativistic~\cite{czachor,harshman_pra05}.

In the case of bi-partite quantum states with finite degrees of freedom, the quantification of the entanglement in a pure state is unambiguous.  Then one may ask, whether there exist S-matrices that are perfect entanglers, i.e. an operator that takes an initially separable state into a maximally entangled state~\cite{zhang02}.  The main result proved here is that rotationally-invariant, perfectly-entangling S-matrices only exist for a particular type of separable initial state.  When they do exist, the scattering phase shifts that determine the S-matrix in the spin sector can be explicitly calculated.  For scattering spin systems like those described above, this paper calculates the in-states and phases necessary for maximal entanglement of spin systems with spins of $\sigma = 1/2$, $1$, and $3/2$.  Reversing the idea, this shows how information about scattering interactions can be gained by looking at the entanglement of the out-state within particular partial waves of orbital angular momentum.

\section{Dynamical entanglement}

Consider two quantum systems with the same finite-number of levels $d$ with Hilbert space $\HS_{d^2} = \HS_d \otimes \HS_d$.  The entropy of entanglement for a pure state $\kt{\psi}\in\HS_{d^2}$ is
\begin{equation}\label{ent}
E(\psi) = S(\rho_1) = S(\rho_2)
\end{equation}
where $\rho_1 = \mathrm{tr}_2 [\kt{\psi}\br{\psi}]$ is the density matrix for system 1 that remains after a partial trace over system 2, and  $S(\rho)= - \mathrm{tr}[\rho\log\rho]$ is the Von Neumann entropy of the density matrix $\rho$.  Conventionally, the logarithm in the entropy is taken in base 2, but for our purposes it is better if it is taken base $d$.  Then the entanglement is bounded by $0 \leq E(\psi) \leq 1$.
A pure state of the form
\begin{eqnarray}\label{noent}
\kt{\phi} &=& \kt{\phi_1} \otimes \kt{\phi_2} \nonumber\\
&= &\sum_{j=0}^{(d-1)} a_j \kt{j} \times \sum_{k=0}^{(d-1)} b_k \kt{k}
\end{eqnarray}
is unentangled. The reduced density matrix is $\rho_i = \kt{\phi_i}\br{\phi_i}$ and $E(\phi) = 0$.
States of the system with the form
\begin{equation}\label{maxent}
\kt{\psi'} = \frac{1}{\sqrt{d}} \sum_{j=0}^{(d-1)} e^{i\alpha_j}\kt{j}\otimes\kt{\pi_j},
\end{equation}
where $\pi_j$ is the $j$-th element of permutation $\pi\in S_d$ of the numbers $\{1,...,d\}$ and $\alpha_j\in\mathbb{R}$, have maximum entanglement $E(\psi) = \log_d d = 1$.  These states have reduced density matrices $\rho_1 = \rho_2 = (1/d) \mathbb{I}_d$. 

How can a maximally-entangled state (\ref{maxent}) evolve from a unentangled pure state (\ref{noent})?  In other words, what operators are perfect entanglers for such a system.  Mathematically, any unit-normalized state in $\HS_{d^2}$ can be transformed to any other unit-normalized state in $\HS_{d^2}$ by unitary transformation $U\in\mathrm{U}(d^2)$.  Since a global phase in not physically meaningful, it is sufficient to consider $\mathrm{SU}(d^2)$.  The subset of local operators $\mathrm{U}(d)\times\mathrm{U}(d)$ cannot change the entropy of entanglement of a pure state, so perfect entanglers must be elements of the set $P_d = \mathrm{SU}(d^2)/\mathrm{U}(d)\times\mathrm{U}(d)$.  Zhang \emph{et al.}~\cite{zhang02} consider the system of two qubits and find that half of $P_2$ are perfect entanglers.

From another perspective, since $\mathrm{U}(N)$ is a connected matrix Lie group, it is possible to express every $U\in\mathrm{U}(d^2)$ as
\begin{equation}\label{un}
U = \exp(iH_1 t_1)\exp(iH_2 t_2)...\exp(iH_m t_m)
\end{equation}
for some finite number of $d^2\times d^2$ Hermitian matrices $\{H_1, H_2,...H_m\}$~\cite{hall}.  In principle, one could imagine some series of interaction Hamiltonians, switched on and off at certain times, that when exponentiated lead to time evolution operators that are perfect entanglers~\cite{zhang02}.  However, in a given physical system, symmetries may limit what kind of Hamiltonians can appear in (\ref{un}), or equivalently many $U\in\mathrm{U}(d^2)$.  

The example we consider in depth here is scattering by central forces of two systems with spin $\sigma$.  Assume that in the limit $t\rightarrow \pm\infty$, the two systems are not interacting.  Then one can define the unitary scattering operator, the S-matrix, that transforms the unentangled in-state $\kt{\phi}$ to the maximally-entangled, out-state $\kt{\phi'}$:
\begin{equation}
\kt{\phi'} = S\kt{\phi}.
\end{equation}
Scattering interactions are spherically symmetric so the S-matrix must commute with the total angular momentum angular momentum operator.  Since it is a central interaction, it must also separately commute with the total spin and orbital angular momentum operators.  The spin sector of the interaction can be separated from the other degrees of freedom and will therefore be identified with $\HS_{d^2}$, where $d=2\sigma +1$, and the rotations $R\in\mathrm{SO}(3)$ will be represented by the tensor product of the single-particle spin representations $D^\sigma(R) \times D^\sigma(R)$.
As a result, the set of possible perfectly-entangling S-matrices is much reduced by the symmetry: they must global, unitary matrices that commutes with every $D^\sigma(R) \times D^\sigma(R)$.  
Additionally, it will be shown that only for a certain set of in-states $\kt{\phi}$ do S-matrices exist that are perfect entanglers and the possibilities can be enumerated.

\section{Useful bases for analyzing spin system dynamical entanglement}

There are two bases that will be used for the states in $\HS_{d^2}$.  The first is the direct product basis denoted by either $\kt{\mu_1 ,\mu_2}$ or $\kt{\chi_1, \chi_2}$.  These are the eigenvectors of individual angular momentum 3-component operators:
\begin{eqnarray}
\Sigma_3^{(1)}\kt{\mu_1 ,\mu_2} &=& \mu_1\kt{\mu_1 ,\mu_2}\nonumber\\
\Sigma_3^{(2)}\kt{\mu_1 ,\mu_2} &=& \mu_2\kt{\mu_1 ,\mu_2}.
\end{eqnarray}
The direct product basis is useful because it is separable like the initially-unentangled state and because measurements are local operations.  Also, the entropy of entanglement requires the partial trace, which is straightforward to evaluate in this basis.

The second useful basis is the direct sum basis $\kt{s\, m}$, which are the eigenvectors of the total angular momentum component operators $\Sigma_3$ and total angular momentum squared operator $\bf{\Sigma}^2$:
\begin{eqnarray}
\Sigma_3\kt{s\, m} &=& m\,\kt{s\, m}\nonumber\\
\bf{\Sigma}^2\kt{s\, m} &=& s(s+1)\,\kt{s\, m}.
\end{eqnarray}
This is called direct sum basis because it arises in the Clebsch-Gordon decomposition of products of irreducible representations (IRs) of the $\SU(2)$ into a direct sum of IRs  It is useful because if the interaction is spherically-symmetric (hence $\SU(2)$-invariant), then the S-matrix has the form
\begin{equation}\label{smat}
\br{s'\, m'}S\kt{s\, m} = e^{2i\delta_s}\delta_{ss'}\delta_{mm'}.
\end{equation}
The convention of calling the $s$-dependent phase $2\delta_s$ comes from scattering theory and the form (\ref{smat}) can be seen as the consequence of the Wigner-Eckhart theorem for scalar operators~\cite{rose} or more generally as the consequence of Schur's lemma~\cite{fuchs}.  In the case of scattering non-relativistic particles, $\delta_s$ also depends on the total angular momentum, orbital angular momentum, and magnitude of relative momentum~\cite{harshman_05a}.  The direct product basis and direct sum basis are connected by the Clebsch-Gordan coefficients (CGCs) for $\SU(2)$, $\bk{s\,m}{\mu_1, \mu_2}$, which can be chosen as real.

So then the question becomes: for what states $\phi$ of the form (\ref{noent}) will $S\phi$ be a maximally entangled state of the form (\ref{maxent}) and what must the phases $\delta_s$ be for this to occur?
To answer this question, the state $\phi' = S\phi$ is expressed in the direct product basis:
\begin{eqnarray}
\kt{\phi'} &=& S\sum_{\mu_1 \mu_2} a_{\mu_1} b_{\mu_2} \kt{\mu_1 ,\mu_2}\nonumber\\
&=& \sum_{\chi_1 \chi_2} c_{\chi_1,\chi_2} \kt{\chi_1,\chi_2}
\end{eqnarray}
where
\begin{equation}
c_{\chi_1\chi_2} = \sum_{\mu_1 \mu_2}a_{\mu_1} b_{\mu_2} r(\chi_1,\chi_2;\mu_1,\mu_2)
\end{equation}
and
\begin{equation}
r(\chi_1,\chi_2;\mu_1,\mu_2) = \sum_{s\,m}\bk{s\,m}{\mu_1 ,\mu_2}\bk{\chi_1 ,\chi_2}{s\,m}e^{2i\delta_s}.
\end{equation}
Using this notation, the reduced density matrix  for particle 1, $\rho_1 = \mathrm{tr}_2(\kb{\phi'}{\phi'})$, is
\begin{equation}\label{red}
\rho_1  = \sum_{\chi_1\chi'_1} \sum_{\chi_2}c_{\chi_1,\chi_2}c^*_{\chi'_1,\chi_2}\kt{\chi_1}\br{\chi'_1}.
\end{equation}
Since $\rho_1 = (1/d)\mathbb{I}_d$ maximizes $S(\rho_1)$, the state $\phi'=S\phi$ will be a maximally entangled state if
\begin{equation}\label{sol}
\sum_{\chi_2}c_{\chi_1,\chi_2}c^*_{\chi'_1,\chi_2} = \frac{1}{d}\delta_{\chi_1,\chi'_1}.
\end{equation}

\section{Finding the in-states and phase shifts}

The equation (\ref{sol}) must be solved to find the possible phase shifts and in-states that lead to maximally-entangled out-states.
The coefficients $c(\chi_1,\chi_2)$ in (\ref{sol}) depend on three things: the initial state $\phi$ through $a_{\mu_1}$ and $b_{\mu_2}$, the CGCs for $\SU(2)$, and the scattering phases $\delta_s$.

The only states for which a perfectly-entangling S-matrix exists, as will be shown below, are states of the form
\begin{equation}\label{instate}
\kt{\phi(u,\lambda)} = U(u)\kt{\lambda,-\lambda},
\end{equation}
where $U(u) = U_1(u)\otimes U_2(u)$ is the direct product representation of the rotation group with $u\in\SU(2)$ and $\lambda\in\{\sigma,\sigma-1,...-\sigma\}$.  Such states are the zero eigenvectors of the total spin operator $\Sigma_3' = (R(u){\bf \Sigma})_3$, where $R(u)$ is the image of $u$ under the standard homomorphism $\SU(2)\rightarrow\mathrm{SO}(3)$.

Because $[S,U(u)]=0$ for all $u\in SU(2)$, we have
\begin{eqnarray}\label{maxout}
\kt{\phi'} &=& U(u)S\kt{\lambda,-\lambda}\nonumber\\
&=& U(u)\sum_s e^{2 i \delta_s}\bk{s\,m=0}{\lambda,-\lambda}\kt{s\, m=0} \nonumber\\
&=& U(u)\sum_{\chi_1\chi_2}\tilde{c}_{\chi_1,\chi_2}(\lambda)\kt{\chi_1, \chi_2},
\end{eqnarray}
where we define
\begin{eqnarray}
\tilde{c}_{\chi_1,\chi_2}(\lambda) &=&  \sum_s e^{2 i \delta_s}\bk{s\,m=0}{\lambda,-\lambda}\bk{\chi_1, \chi_2}{s\,m=0}\nonumber\\
& =& g_{\chi_1}(\lambda) \delta_{\chi_1,-\chi_2}.
\end{eqnarray}
Then (\ref{red}) becomes
\begin{equation}\label{redlam}
\rho_1 = U_1(u)\left(\sum_{\chi_1} |g_{\chi_1}(\lambda)|^2 \kb{\chi_1}{\chi_1}\right) U^\dag_1(u).
\end{equation}
This is a diagonal matrix and $|g_{\chi_1}(\lambda)|^2$ are the Schmidt coefficients for the state after the interaction.  If all the $\delta_s$ are the same phase, then   $|g_{\chi_1}(\lambda)|^2=\delta_{\chi_1,\lambda}$ because the basis transformation given by Clebsch-Gordan decomposition is unitary.  In this case, the dynamics just evolve the in-state by a total phase and there is no entanglement.  The reduced density matrix $\rho_1$ in (\ref{redlam}) will be of the maximally entangled form for a given $\lambda$ if and only if
\begin{equation}\label{cond}
|g_{\chi_1}(\lambda)|^2 = 1/d\ \mbox{for all}\ \chi_1.
\end{equation}
If (\ref{cond}) is satisfied, then the density matrix will be a scalar multiple of the identity and commute with all rotations $u\in\SU(2)$.  It can be shown that (\ref{redlam}) would not be diagonal for any other initial condition besides one of the form $\phi (u,\lambda)$.  Only eigenvectors of the total angular momentum component (in any direction) with $m=0$ lead to a reduced density matrix of the form (\ref{redlam}) because only those states can have non-zero Clebsch-Gordan coefficients with every $s$ from zero to $2\sigma$.
Also, this shows that all maximally-entangled states that emerge from a scattering experiment will have the form in (\ref{maxout}).

When the in-state $kt{\phi(u,\lambda)}$ has the form (\ref{instate}), then 
the set of $\delta_s$ which satisfy of $|g_{\chi_1}(\lambda)|^2 = 1/d$ for all $\chi_1\in\{\sigma,\sigma-1,...-\sigma\}$, will determine a perfectly-entangling S-matrix.Explicit solutions have been found for $\sigma = 0$ (trivial), $1/2$, $1$, and $3/2$ and it has been found that the solutions are in fact independent of $\lambda$ and only depend on $\sigma$.  The results are summarized below.  The phases are set so $\delta_0=0$ and all other phases are relative to this and all $\delta_s\in(-\pi,\pi]$.
\begin{itemize}
\item For $\sigma=0$, spin entanglement is not meaningful.
\item For $\sigma=1/2$, (\ref{cond}) for all $\lambda$ and $\chi_1$ leads to two independent equations:
\begin{eqnarray*}
\frac{1}{2}&=&\cos^2\delta_1\\
\frac{1}{2}&=&\sin^2\delta_1.
\end{eqnarray*}
By fixing the global phase so that $\delta_0=0$, the geometric structure of possible S-matrices $[\delta_1]$ is isomorphic to the circle $S^1$ and the four perfectly-entangling S-matrices are the points $\delta_1=\pm 3\pi/4$ or $\delta_1=\pm\pi/4$.
\item For $\sigma=1$, (\ref{cond}) for all $\lambda$ and $\chi_1$ leads to three independent equations:
\begin{eqnarray*}
\frac{1}{3} &=& \frac{1}{18}\left(7 + 6\cos(2\delta_1) +  3\cos(2\delta_1 - 2\delta_2)+\cos(2\delta_2)\right)
\\
\frac{1}{3} &=& \frac{4}{9}\sin^2\delta_2\\
\frac{1}{3} &=& \frac{1}{18}\left(7 - 6\cos(2\delta_1)- 3\cos(2\delta_1 - 2\delta_2)+2\cos(2\delta_2)\right).
\end{eqnarray*}
With fixed $\delta_0$, the geometric structure of possible S-matrices $[\delta_1, \delta_2]$ is isomorphic to the torus $T^2 = S^1 \times S^1$ and the eight perfectly-entangling S-matrices are the points $\{\delta_1,\delta_2\}=\{\pi/12 \pm \pi/4, - \pi/6 \pm \pi/2\},\{-\pi/12 \pm \pi/4, \pi/6 \pm \pi/2\}$.
\item For $\sigma=3/2$, (\ref{cond}) leads to four independent equations:
\begin{eqnarray*}
\frac{1}{4} &=& \frac{1}{400}\left(132 + 90\cos(2\delta_1) +  90\cos(2\delta_1 - 2\delta_2)+ 18\cos(2\delta_1 - 2\delta_3) \right.\\
&&\left.+  50\cos(2\delta_2) + 10\cos(2\delta_2 - 2\delta_3) +  10\cos(2\delta_3) \right)
\\
\frac{1}{4} &=& \frac{1}{400}\left(132 - 90\cos(2\delta_1) -  90\cos(2\delta_1 - 2\delta_2)+ 18\cos(2\delta_1 - 2\delta_3) \right.\\
&&\left.+  50\cos(2\delta_2) - 10\cos(2\delta_2 - 2\delta_3) -  10\cos(2\delta_3) \right)
\\
\frac{1}{4} &=& \frac{1}{400}\left(68 - 30\cos(2\delta_1) +  30\cos(2\delta_1 - 2\delta_2)- 18\cos(2\delta_1 - 2\delta_3) \right.\\
&&\left.-  50\cos(2\delta_2) - 30\cos(2\delta_2 - 2\delta_3) +  30\cos(2\delta_3) \right)
\\
\frac{1}{4} &=& \frac{1}{400}\left(68 + 30\cos(2\delta_1) -  30\cos(2\delta_1 - 2\delta_2)- 18\cos(2\delta_1 - 2\delta_3) \right.\\
&&\left.-  50\cos(2\delta_2) + 30\cos(2\delta_2 - 2\delta_3) -  30\cos(2\delta_3) \right)
\end{eqnarray*}
With fixed $\delta_0$, the geometric structure of possible S-matrices $[\delta_1, \delta_2, \delta_3]$ is isomorphic to the 3-torus $T^3 = S^1 \times S^1\times S^1$ and there are four sets of solutions for any value of $\delta_1\in(-\pi,\pi]$: $\delta_2 = \pm\pi/2$ and $\delta_3 = \delta_1$ or $\delta_2 = \pm\pi/2$ and $\delta_3 = \delta_1 \pm \pi$ (whichever one of $\delta_1 \pm \pi\in(-\pi,\pi]$).
\end{itemize}
Solutions for $\sigma \geq 2$ have not yet been investigated.

\section{Conclusion}

In summary, it has been shown that a rotational-symmetric interaction acts as a constraint on dynamical generation of entanglement in scattering systems.  Only for certain initial states do there exist perfectly-entangling S-matrices, only certain phase shifts allow for such a transformation, and only a subset of maximally entangled vectors emerge as out-states.  The implications for production of entangled states by scattering-type interactions are worth exploration and could guide experimentalists in constructing or searching for suitable systems and for tuning the interactions of those systems where possible.  Additionally, this idea can be reversed, and as in classic partial wave analysis, entanglement correlations could be used to find information on the phase shifts.  This idea has been partially explored in \cite{kurizki05} for translational entanglement in scattering, but much work remains to be done.  

Finally, this paper can also be thought of as showing how interaction symmetries limit the possible unitary transformations and therefore limit the maximal entanglement possible for a given initial condition. Particles are elements of the representation spaces of space-time symmetry groups.  When direct products of these representation spaces are decomposed into a direct sum of irreducible representations, one can represent the S-matrix in a basis that makes global symmetries explicit.  In this case, it has been shown that for unitary irreducible representations of the Galilean group, the spin degrees of freedom and be separated from the translational degrees of freedom and the consequences of global rotational symmetry can be explored.  Extensions to other space-time and interaction symmetry groups will be considered in the future.

\section*{Acknowledgments}
The author would like to thank Mark Byrd for fruitful discussions.

\end{document}